\documentclass[11pt]{article}
\usepackage{amsmath,graphicx,authblk,setspace}
\usepackage{booktabs} 
\usepackage[numbers]{natbib}
\usepackage[colorlinks=true,citecolor=blue,linkcolor=blue,urlcolor=blue]{hyperref}
\usepackage[margin=0.9in]{geometry}
\usepackage{booktabs}
\usepackage{amsmath}
\usepackage{threeparttable}
\usepackage{chngcntr}

\renewcommand\thetable{\arabic{table}}
\title{Quantifying Diagnostic Signal Decay in Dementia: A National Study of Medicare Hospitalization Data}
\author[1,2]{Federica Spoto}
\author[3]{Jiazi Tian}
\author[3,4,5]{Jonas Hügel}
\author[4]{Daniel T. Ortega}
\author[3]{Christine S. Ritchie}
\author[6,7]{Deborah Blacker}
\author[1]{Francesca Dominici}
\author[8]{Chirag J. Patel}
\author[1]{Daniel Mork}
\author[3]{Hossein Estiri\thanks{To whom correspondence should be addressed; E-mail: HESTIRI@mgh.harvard.edu}}

\affil[1]{Department of Biostatistics, Harvard T.H. Chan School of Public Health, Boston, MA, USA}
\affil[2]{Department of Environmental Health, Harvard T.H. Chan School of Public Health, Boston, MA, USA}
\affil[3]{Department of Medicine, Massachusetts General Brigham, Boston, MA, USA}
\affil[4]{Department of Medical Informatics, University Medical Center Göttingen, Göttingen, Germany}
\affil[5]{Campus Institute Data Science, Section of Medical Data Science, University of Göttingen, Göttingen, Germany}
\affil[6]{Department of Psychiatry, Massachusetts General Hospital, Boston, MA, USA}
\affil[7]{Department of Epidemiology, Harvard T.H. Chan School of Public Health, Boston, MA, USA}
\affil[8]{Department of Biomedical Informatics, Harvard Medical School, Boston, MA, USA}
\date{}

\begin{document}

\maketitle
\vspace{-1cm}
\abstract{\textbf{Background:} Artificial intelligence (AI) models in healthcare depend on the fidelity of diagnostic data, yet the quality of such data is often compromised by variability in clinical documentation practices. In dementia, a condition already prone to diagnostic ambiguity, this variability may introduce systematic distortion into claims-based research and AI model development.

\textbf{Methods:} We analyzed Medicare Part A hospitalization data from 2016–2018 to examine patterns of dementia-related ICD-10 code utilization across more than 3,000 U.S. counties. Using a clinically informed classification of 17 ICD-10 codes grouped into five diagnostic categories, we applied the transitive Sequential Pattern Mining (tSPM+) algorithm to model temporal usage structures. We then used matrix similarity methods to compare local diagnostic patterns to national norms and fit multivariable linear regressions to identify county-level demographic and structural correlates of divergence.
 
\textbf{Findings:} We found substantial geographic and demographic variation in dementia-related diagnostic code usage. Non-specific codes were dominant nationwide, while Alzheimer’s disease and vascular dementia codes showed pronounced variability. Temporal sequence analysis revealed consistent transitions from specific to non-specific codes, which suggest degradation of diagnostic specificity over time. Counties with higher proportions of rural residents, Medicaid-eligible patients, and Black or Hispanic dementia patients demonstrated significantly lower similarity to national usage patterns. Our model explained 38\% of the variation in local-to-national diagnostic alignment. 

\textbf{Interpretation:} These findings reveal widespread and structured inconsistencies in dementia documentation across the U.S. healthcare system. While some of this variation may reflect genuine differences in dementia incidence, prevalence, and population-level care-seeking behavior or access to diagnostic services, a substantial portion is likely attributable to differences in diagnostic practices and documentation. Much of this can be characterized as diagnostic signal decay, a progressive loss or distortion of clinical meaning as data pass through heterogeneous healthcare systems. Such decay poses a foundational challenge for AI development and claims-based research by introducing context-dependent bias into algorithmic pipelines. Future work could explore developing an index to quantify diagnostic data quality and guide model calibration and deployment. Accounting for diagnostic fidelity, including through linkage with systematically evaluated populations such as the Health and Retirement Study (HRS), where sources of variability can be more precisely disentangled, will be essential to ensure the accuracy, fairness, and generalizability of data-driven tools in dementia care.}

\section{Introduction}

Artificial intelligence (AI) and machine learning models are rapidly reshaping how clinicians identify, monitor, and manage neurodegenerative conditions like Alzheimer’s disease and related dementias (ADRD)\citep{mmadumbu2025early, habbal2025harnessing, hao2024early, sadeghian2024methods, kale2024ai}. These models increasingly rely on administrative and electronic health record (EHR) data (particularly diagnostic codes) to define phenotypes, identify at-risk cohorts, and predict outcomes. While the computational capacity to detect complex patterns has grown, the fidelity of the clinical signals embedded in these data has received far less scrutiny. Nowhere is this gap more consequential than in dementia, where diagnosis is often missing, delayed, ambiguous, or inconsistently documented across healthcare settings \citep{bynum2024regional, dieleman2025drivers, wennberg1999understanding}.
We hypothesize that these inconsistencies are manifestations of a broader phenomenon we term diagnostic signal decay. Unlike biomarker assays or structured imaging reports, diagnosis codes are not direct reflections of pathophysiology. They are artifacts of a sociotechnical process shaped by institutional policies, provider incentives, EHR interfaces, documentation culture, health care systems dynamics \citep{agniel2018biases}, access to care, and social context \citep{dieleman2025drivers, MedPAC2024, li2021elucidating}. This complex and often fragmented pipeline introduces semantic drift, where the same ICD-10 code may carry different meanings depending on clinical context, practice environment, or region. Over time and across populations, this contributes to “signal decay”: a degradation in the fidelity with which clinical reality is captured in coded data.

In dementia, underrecognition remains a more pervasive challenge than diagnostic ambiguity. Its onset is often insidious, subtle, and easily overlooked by both clinicians and patients. Although the syndrome eventually manifests with protean features and diagnostic uncertainty, particularly in later stages, the frequent use of vague or non-specific codes (e.g., “unspecified dementia”) reflects not just the complexity of the disease, but also the systemic constraints of clinical documentation, especially in resource-limited settings \citep{bynum2024regional, dieleman2025drivers, mainor2019icd}. Such practices obscure clinical subtypes, delay appropriate interventions, and distort both epidemiologic surveillance and AI model training. The consequence is a dual risk: the erosion of data quality and the amplification of health disparities through models that fail to generalize across structurally marginalized populations and fail to provide diagnostic specificity \citep{xue2024ai, lee2024robust, kornblith2022association, katz2012age, mehta2017systematic, fitzpatrick2004incidence, plassman2007prevalence}.

Although validated phenotyping algorithms have been developed to improve the consistency and validity of ADRD identification in administrative data \citep{mccarthy2022validation, ccw}, these tools remain susceptible to the upstream variability in diagnostic code usage. Previous work has identified racial and geographic disparities in dementia incidence, outcomes, and access to care \cite{moon2019dementia, mayeda2016inequalities, hayward2021importance, ornstein2018medicare, gorges2019national, gessert2006rural}, but few studies have evaluated how regional and population-level variation in diagnostic coding practices may systematically alter the informational content of datasets used to train predictive models.

In this study, we analyze Medicare inpatient claims from 2016–2018 to assess the magnitude and drivers of diagnostic signal decay in dementia-related ICD-10 code utilization. We apply a clinically informed classification of 17 ICD-10 codes, grouped into five diagnostic categories, and use transitive Sequential Pattern Mining (tSPM+) \citep{estiri2020transitive, hugel2023tspm+,hugel2023tspmplus} to uncover the temporal structure of code usage across over 3,000 U.S. counties. We then use matrix similarity methods to compare local versus national utilization patterns and multivariable regression to identify the demographic and socioeconomic correlates of diagnostic divergence.

By quantifying geographic and racial heterogeneity in dementia diagnostic coding, we provide a framework for understanding and potentially correcting for diagnostic signal decay in real-world data. These insights are essential not only for improving the internal validity of health services research but also for ensuring that AI models trained on such data do not inadvertently propagate structural inequities. Addressing diagnostic signal decay is thus a foundational step toward building robust and generalizable data-driven tools for dementia care.

\section{Background}

Dementia disproportionately affects structurally marginalized populations in the United States. Black and Hispanic individuals experience higher incidence and are more likely to be hospitalized while living with dementia; a pattern that likely reflects not only differences in disease burden but also broader disparities in access to outpatient support, comorbid condition management, and alternatives to hospital-based care, which persist beyond clinical risk factors and mirror underlying social and environmental inequities \citep{kornblith2022association, katz2012age, mehta2017systematic, fitzpatrick2004incidence, plassman2007prevalence, moon2019dementia, mayeda2016inequalities, hayward2021importance, zhu2021sex}. Yet superimposed on these inequities is a less visible but equally critical issue: variability in how dementia is diagnosed and recorded in clinical data.

Across U.S. regions, diagnostic coding practices differ markedly, influenced by institutional incentives, provider norms, and healthcare fragmentation \citep{MedPAC2024, dartmouth1996atlas, smith2011dartmouth, badinski2023geographic}. The adoption of ICD-10 in 2015 enabled greater specificity but introduced discontinuities that hinder comparison across time and space \citep{mainor2019icd}. Practices such as upcoding, diagnostic ambiguity, and regional “DRG creep” further erode the interpretability of claims data \citep{marshall2022diagnosis, geruso2020upcoding, crespin2024upcoding, carter1990much, drgcreep1981}. In dementia care, where diagnosis is inherently complex, these issues manifest as diagnostic signal decay: a systematic degradation of data fidelity caused by inconsistencies in how clinical reality is translated into administrative codes.

Non-specific dementia codes (e.g., F03) are commonly used, particularly in settings where diagnostic evaluations by cognitive specialists are limited. While this pattern is especially pronounced in inpatient data, where coding is often performed by professional coders relying on pre-existing chart information, the issue is not confined to hospitals. In outpatient settings, primary care physicians, who may be the first to document cognitive concerns, frequently opt for vague terms like “memory loss” or “memory problems”. This can reflect time constraints, limited diagnostic certainty, or a desire to avoid stigmatizing labels. Such nonspecific coding choices are often perpetuated downstream by other providers or coders not focused on cognition. The result is a proliferation of broad, non-specific diagnostic codes that obscure true disease subtypes, limit the fidelity of claims-based research, and compromise the generalizability of AI models designed for dementia detection and stratification.. Despite validated case ascertainment algorithms \citep{mccarthy2022validation,ccw}, few studies have quantified how this diagnostic distortion varies geographically or demographically, or how it impacts algorithmic fairness.

By interrogating the structure and drivers of diagnostic signal decay in dementia coding, we can illuminate how inequities in care are encoded into data, and how they might be corrected in the age of computational medicine.

\section{Methodology}

\subsection{Study Framework and Hypotheses}
This study was designed to empirically assess “diagnostic signal decay”, defined here as the degradation of informational fidelity in administrative claims due to institutional, demographic, and regional variability in coding practices. We conceptualize this decay as a structured process influenced by health system fragmentation, provider behavior, and differential access to diagnostic precision.

We tested two primary hypotheses:

\begin{itemize}
  \item[H1](Signal Variability Hypothesis): Dementia-related ICD-10 code utilization exhibits significant variation across racial, ethnic, and geographic strata, reflecting differential signal fidelity in how dementia is recorded.
  \item[H2] (Structural Signal Decay Hypothesis): Geographic deviations from nationally observed diagnostic patterns are systematically associated with county-level sociodemographic and structural characteristics, indicating that signal decay is not random, but shaped by structural factors such as poverty, healthcare access, and racial composition.
\end{itemize}

To test these hypotheses, we developed a computational framework to measure and interpret dementia-related diagnostic signal patterns at both the national and local levels.

\subsection{Data}
We obtained Medicare Part A MedPAR data files from the Centers for Medicare and Medicaid Services (CMS), encompassing all inpatient hospitalization claims for Medicare fee-for-service (FFS) beneficiaries aged 65 years and older between January 1, 2016, and December 31, 2018. Each record contained demographic information, including age, sex, race, ethnicity, Medicaid dual-eligibility status, and residential zip code. Additionally, the dataset provided information regarding the original reason for Medicare eligibility—whether due to age, disability insurance benefits, or end-stage renal disease. Each beneficiary was assigned a unique identifier to facilitate the longitudinal tracking of hospitalization claims across the study period.

Each hospitalization claim included up to 25 billing diagnoses based on the International Classification of Diseases, Tenth Revision (ICD-10). We identified a set of 17 ICD-10 codes indicative of dementia, designated as dementia-qualifying codes. The selected codes were grouped into five diagnostic categories: vascular dementia, non-specific dementia, Alzheimer’s disease, Pick’s disease (now more commonly called behavioral variant frontotemporal dementia (bvFTD)), and other neurocognitive disorders (see Table \ref{tab:codes} in Supplementary Materials).

We included in our analysis all beneficiaries who experienced at least one hospitalization with a dementia-qualifying ICD-10 code during the study period and who resided in the same zip code continuously throughout. This criterion was applied to ensure that measures reflecting the HUP of a given geographic area accurately represented the population residing there. We restricted our analytic sample to hospitalizations where the source of admission was not a nursing facility and where at least one dementia-qualifying code was present in the billing documentation.

Additionally, we considered county-level data on educational attainment, unemployment rates, and rural population proportions, obtained from the IPUMS National Historical Geographic Information System (NHGIS)\citep{Manson2021}. Specifically, the percentage of individuals with less than a high school education and the unemployment rate were derived from summary tables based on the 2016–2020 American Community Survey (ACS) 5-Year Estimates. The proportion of rural residents was based on data from the 2020 Decennial Census, using the U.S. Census Bureau’s 2020 definition of rural, which classifies all population located outside of urban areas (defined as areas with at least 5,000 residents or 2,000 housing units) as rural.

\subsection{Capturing Diagnostic Signal Structure}
We operationalized diagnostic signal fidelity using both frequency and temporal association structures:

\begin{enumerate}\renewcommand{\labelenumi}{\alph{enumi})}
  \item Descriptive Code Frequency: We quantified the prevalence of each ICD-10 code and diagnostic category nationally and across subgroups to characterize baseline variability.
  
  \item Temporal Signal Structure (tSPM+ Algorithm): To capture the temporal architecture of dementia coding, we applied the transitive Sequential Pattern Mining algorithm (tSPM+)\citep{estiri2020transitive, hugel2023tspm+,hugel2023tspmplus}, which computes patient-level sequences of diagnosis code occurrences and their transitions. To quantify the strength and direction of temporal associations between the usage of the ICD-10 code within each sequence, we compute the Spearman rank-order correlations and p-values with the Holm–Bonferroni adjustment for multiple comparisons to measure temporal associations rho ($\rho$) between diagnosis in the four temporal buckets, representing short-term associations (co-occurence, and association within 1-3 months) and longer-term associations ($>$3 months). 
  \item Signal Similarity Assessment: We compared correlation matrices between subgroups and the national reference using the random skewers method (RS) \citep{cheverud2007research, rohlf2017method}. RS is a matrix similarity approach that quantifies divergence in temporal association structures. A similarity score approaching 1 indicates high fidelity (low decay), while lower scores suggest structural divergence from national patterns.
\end{enumerate}

\subsection{Modeling Drivers of Signal Decay}

To test whether structural and demographic factors predict deviations in diagnostic signal fidelity, we fit a multivariable linear regression model:

\begin{enumerate}
  \renewcommand{\labelenumi}{}
  \item Outcome: Similarity score between each county’s dementia diagnostic pattern and the national reference (as computed by random skewers). Figure \ref{fig:figS1} in Supplementary Materials shows the distribution of the similarity score.
  \item Predictors: County-level indicators of education, unemployment, rurality, Medicaid eligibility, disability, racial/ethnic composition, and overall dementia prevalence. State-level fixed effects were included to account for policy heterogeneity.
\end{enumerate}
This model quantifies the extent to which structural characteristics explain observed signal decay – i.e., the diagnostic drift from national norms. Full description of study variables and statistical modeling is presented in the Supplementary document.

\section{Results}
Our final analytic sample included 1,827,468 Medicare fee-for-service beneficiaries aged $\ge$65 years with at least one hospitalization for which a dementia-related ICD-10 diagnosis was coded between 2016 and 2018, totaling 3,045,819 hospitalizations. The mean age at first qualifying hospitalization was 84 years (SD 7.8), and 60.4\% of patients were female. Most individuals were non-Hispanic White (79.8\%), with 32.7\% eligible for Medicaid at the time of hospitalization (Table ~\ref{tab:table1}).

\begin{table}[ht]
\centering
\begin{tabular}{ll}
\toprule
\multicolumn{2}{l}{\textbf{Study cohort}} \\
\midrule
\quad Number of individuals      & 1,827,468 \\
\quad Number of hospitalizations & 3,045,819 \\
\addlinespace[0.5em]
\midrule
\multicolumn{2}{l}{\textbf{Patient characteristics, N (\%)}} \\
\midrule
\quad \textit{Sex}              & \\
\qquad Male                     & 722,540 (39.5\%) \\
\qquad Female                   & 1,104,012 (60.4\%) \\
\qquad Unknown                  & 916 (0.1\%) \\
\quad \textit{Age}              & \\
\qquad 65--69                   & 94,928 (5.2\%) \\
\qquad 70--74                   & 165,662 (9.1\%) \\
\qquad 75--79                   & 264,366 (14.5\%) \\
\qquad 80--84                   & 376,377 (20.6\%) \\
\qquad 85--89                   & 449,372 (24.6\%) \\
\qquad 90--94                   & 345,317 (18.9\%) \\
\qquad 95+                      & 131,446 (7.2\%) \\
\quad \textit{Race}             & \\
\qquad Hispanic                 & 105,189 (5.8\%) \\
\qquad Native American and Alaska Native & 8,420 (0.5\%) \\
\qquad Non-Hispanic Asian       & 39,414 (2.2\%) \\
\qquad Non-Hispanic Black       & 194,778 (10.7\%) \\
\qquad Non-Hispanic White       & 1,458,488 (79.8\%) \\
\qquad Other                    & 10,141 (0.6\%) \\
\qquad Unknown                  & 11,038 (0.6\%) \\
\quad \textit{Medicaid eligibility} & \\
\qquad Ineligible               & 1,230,077 (67.3\%) \\
\qquad Eligible                 & 597,391 (32.7\%) \\
\bottomrule
\end{tabular}
\caption{Characteristics of Medicare beneficiaries aged 65 years and older who had a dementia-related hospitalization between 2016 and 2018 across the US.}
\label{tab:table1}
\end{table}

\subsection{Prevalence of Diagnostic Categories: Frequency as Signal Density}
\begin{figure}[ht!]
    \centering
    \includegraphics[width=0.8\linewidth]{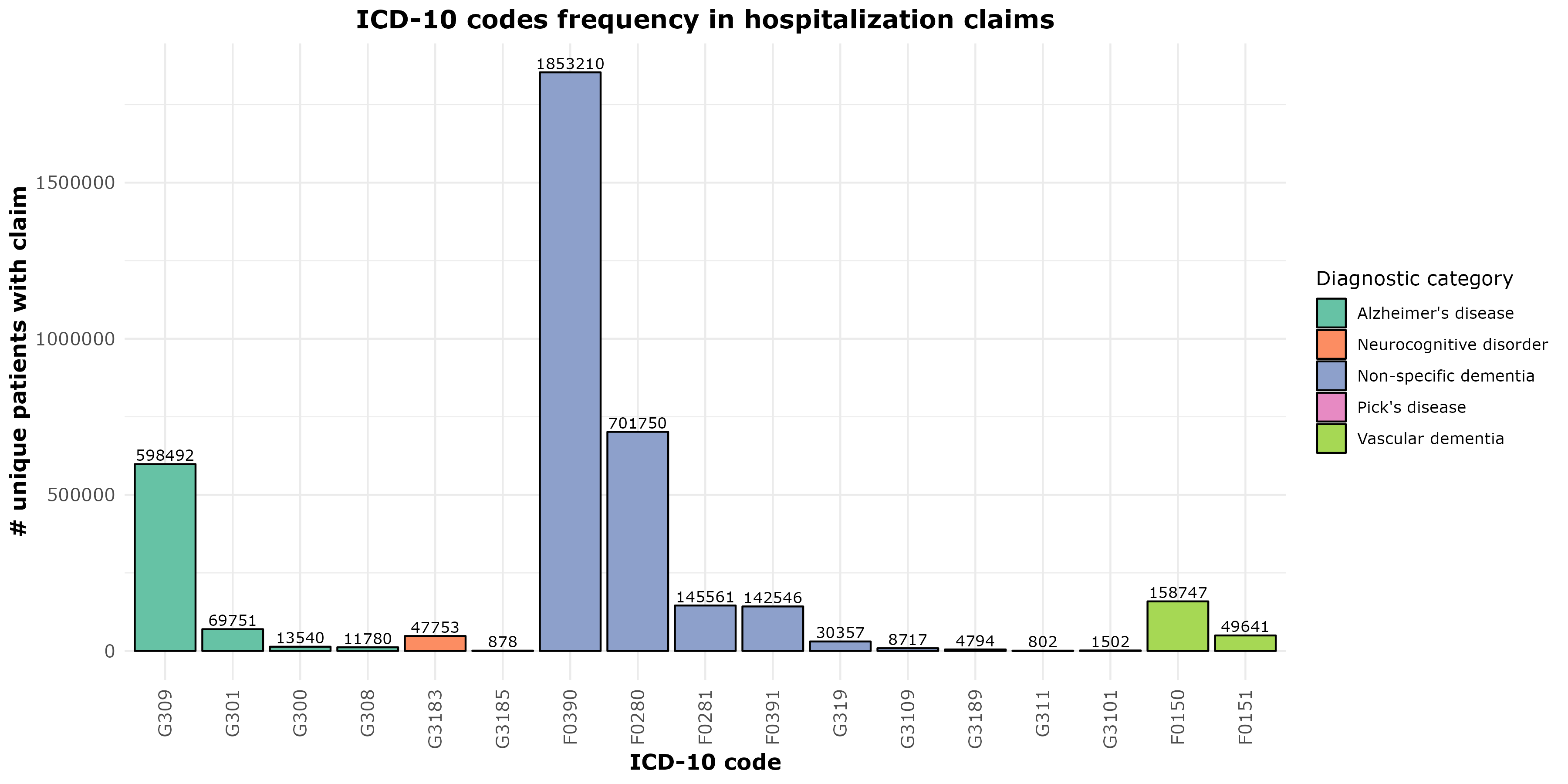}
    \caption{\textbf{Frequency of ICD-10 codes.} The plot shows, for each dementia-qualifying ICD-10 code, how many patients had at least one hospitalization with that code across the contiguous US from 2016 to 2018.}
    \label{fig:fig1}
\end{figure}
\begin{figure}[ht!]
    \centering
    \includegraphics[width=0.9\linewidth]{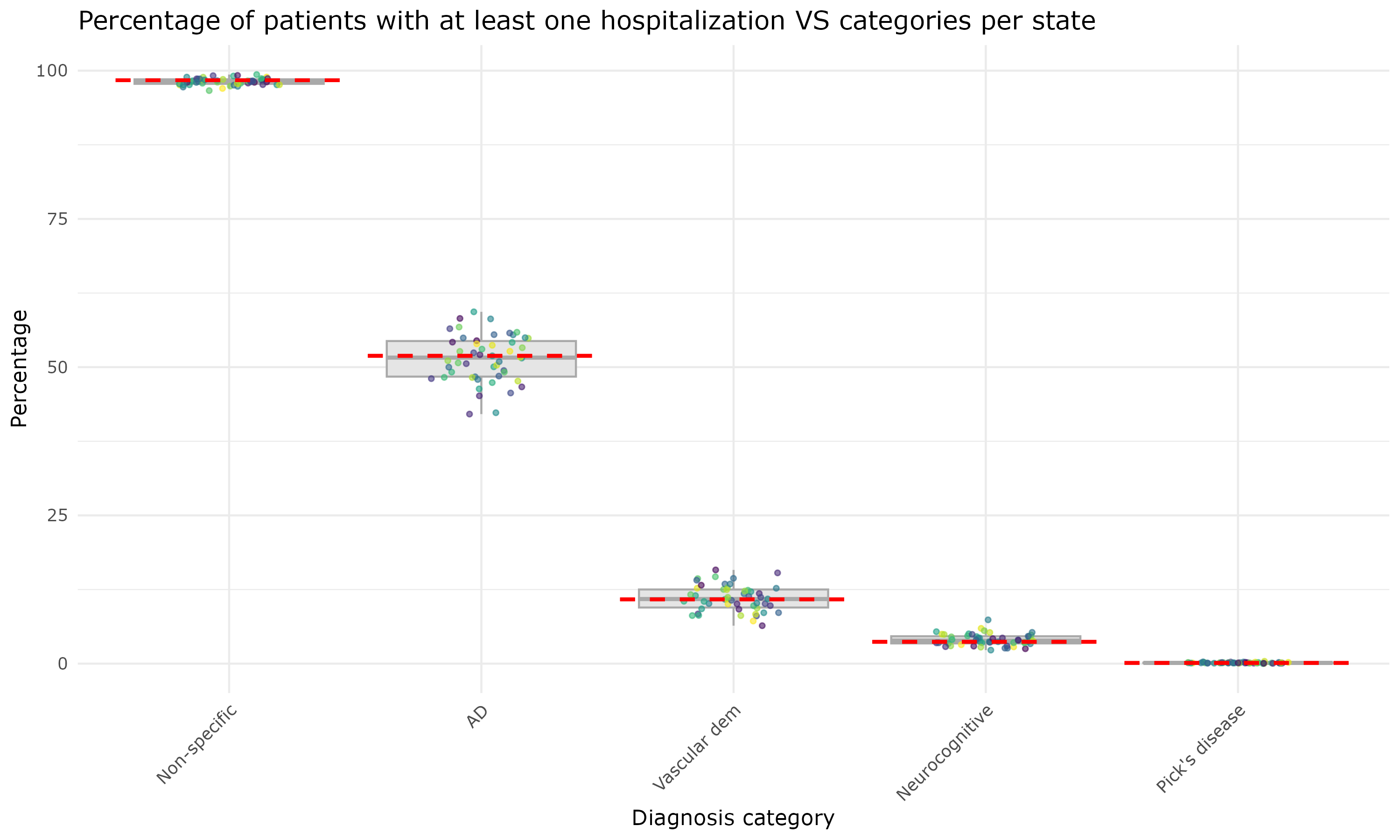}
    \caption{\textbf{State percentage of diagnosis category compared to the nation.} The plot shows, for each diagnosis category, the percentage of patients who had at least one hospitalization with an ICD-10 code in that category across states. The red dashed lines represent the national counterparts.}
    \label{fig:fig2}
\end{figure}

Non-specific dementia codes were the most frequently used, with ICD-10 code F0390 (“Unspecified dementia, unspecified severity”) representing the single most common diagnosis nationally (Figure \ref{fig:fig1}). Alzheimer’s disease codes (e.g., G309) and vascular dementia codes (e.g., F0150) exhibited more limited but variable usage. The predominance of non-specific codes reflects reduced diagnostic resolution, an early indicator of signal density loss in administrative data. This pattern persisted across all U.S. states, with non-specific codes appearing in over 75\% of dementia hospitalizations nationally (Figure \ref{fig:fig2}).

As described in the methods section, to minimize capturing dismissible variability in ICD-10 codes, we grouped the 17 codes into 5 categories – see Table S1 in the appendix.  Analyzing variations in the usage of diagnosis codes (Figure \ref{fig:fig2}) showed a consistently high utilization of the non-specific dementia group of dementia ICD-10 codes (including F0280, F0281, F0390, F0391, G3109, G311, G3189, G319). The highest variable utilization was in the Alzheimer's disease group (including G300, G301, G308, G309) ranging between 42.1\% and 59.3\%, with national average 51.9\%, followed by Vascular dementia codes (including F0150, F0151) ranging between 6.41\% and 15.8\%, with national average 10.8\%. 

\subsection{Temporal Signal Architecture: Code Transition Patterns}

\begin{figure}[ht!]
    \centering
    \includegraphics[width=0.8\linewidth]{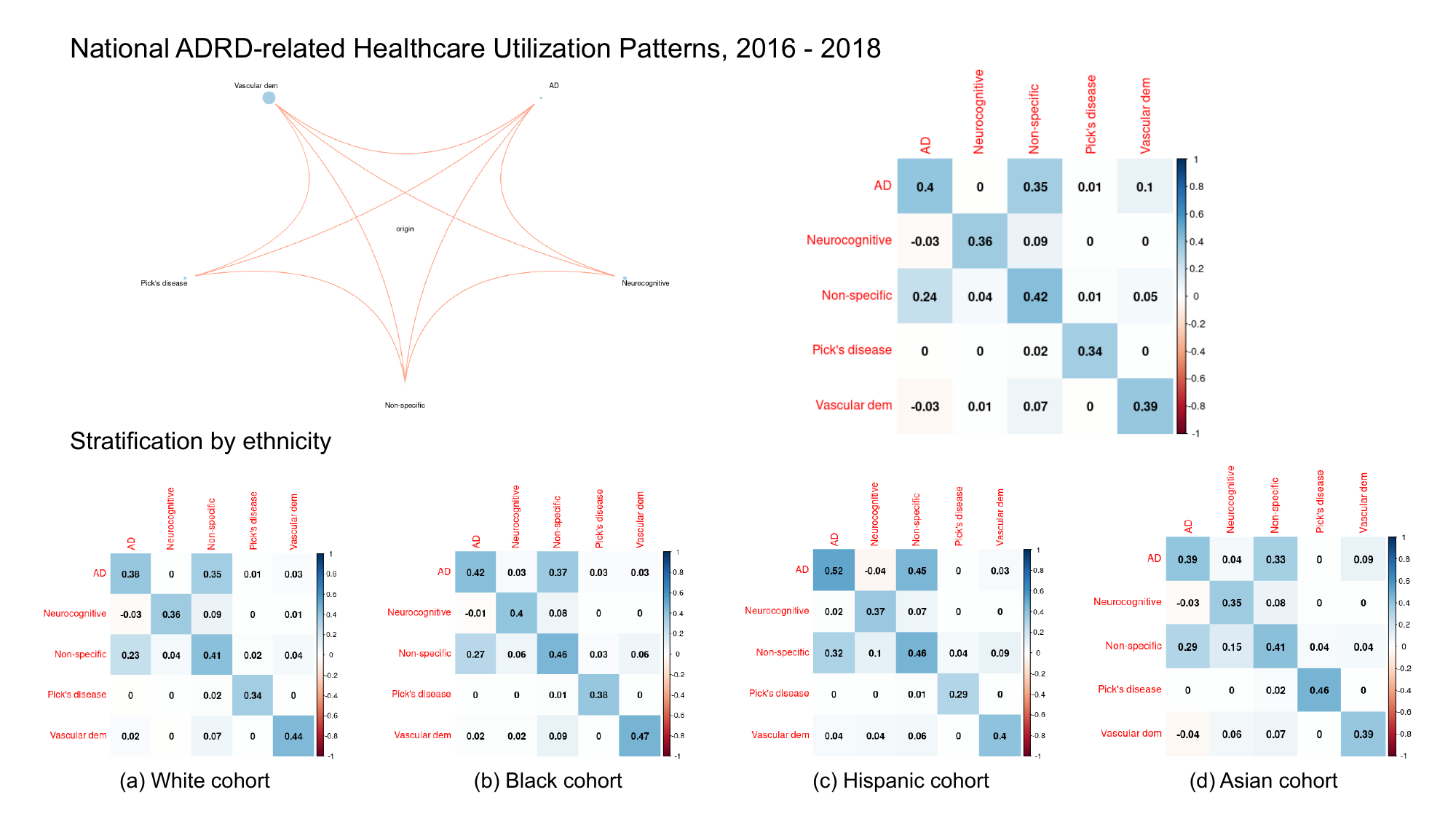}
    \caption{\textbf{tSPM+ algorithm output.} The plot shows the Spearman correlation obtained by running the tSPM+ algorithm. The graph nodes represent the diagnosis categories, whose dimension represents the utilization frequency, while the edge shows the presence of a significant correlation for the sequences given by pairs of ICD-10 codes within different diagnosis categories. The upper part of the figure shows the output when considering the entire cohort and how the correlations were rearranged in matrix form. The lower part shows the different HUP stratifying the cohort by ethnicity.}
    \label{fig:fig3}
\end{figure}

Application of the tSPM+ algorithm revealed that temporal correlations were strongest within diagnostic categories, particularly for non-specific dementia (Figure \ref{fig:fig3}). These self-associations suggest persistent documentation of vague diagnoses over time—e.g., patients initially coded with non-specific dementia were likely to receive similar codes in subsequent hospitalizations. This pattern may, in part, reflect the design of electronic health record systems such as Epic, which are configured to facilitate the carry-forward of prior codes, making it the path of least resistance for both clinicians and medical coders.

Transitions from specific to non-specific codes, especially from Alzheimer’s disease to non-specific dementia, were consistently observed, further supporting the concept of temporal signal decay, where the granularity of diagnosis erodes longitudinally. Stratified analyses by race/ethnicity revealed minimal deviation in overall temporal structure, with Random Skewers (RS) similarity values of 0.98–0.99 across groups, indicating high alignment with national-level signal architecture.

\subsection{Geographic Signal Divergence}
\begin{figure}[ht!]
    \centering
    \includegraphics[width=0.9\linewidth]{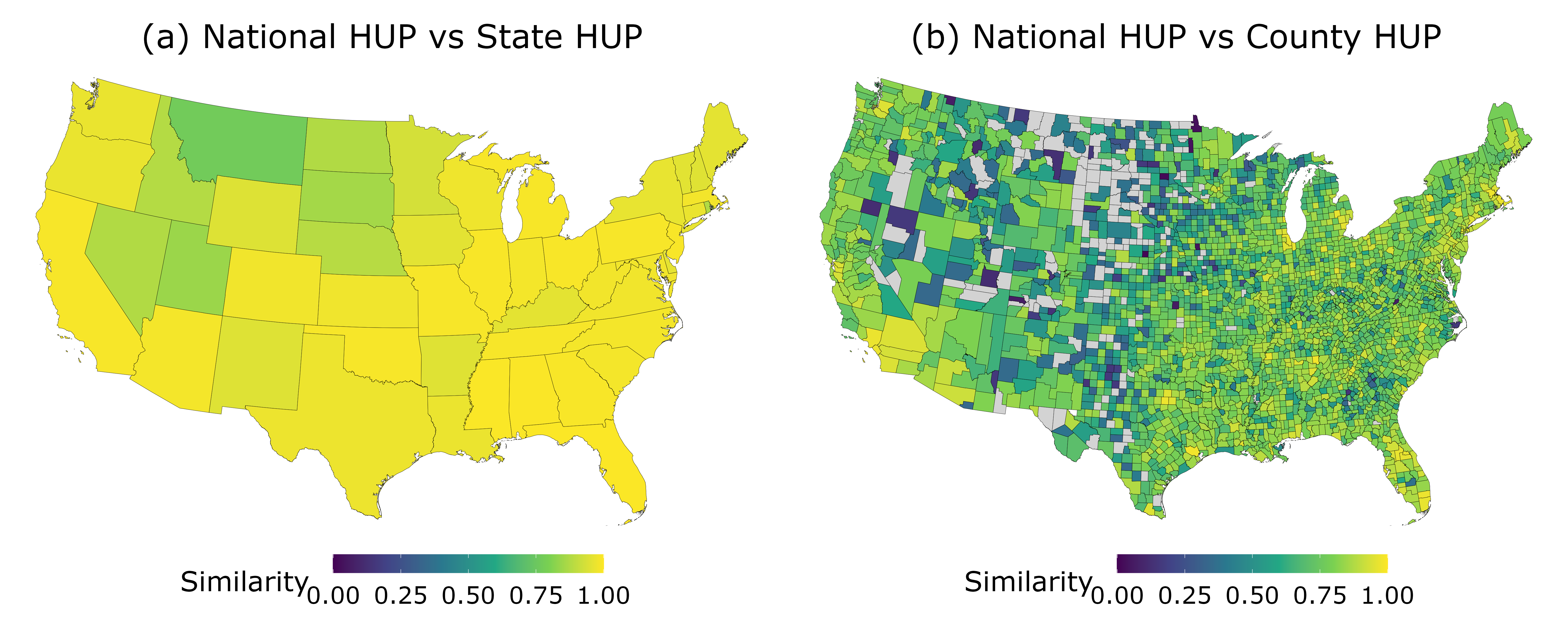}
    \caption{\textbf{Similarity between national and local usage with random skewers.} Panel (a) shows the similarity between national and state usage, while panel (b) compares national and county usage. Counties with fewer than 11 patients with dementia-related hospitalizations were censored and are shown in grey.}
    \label{fig:fig4}
\end{figure}

Despite overall consistency in racial subgroup patterns, we observed substantial geographic variation in diagnostic code utilization structures. Using the random skewers method, we computed matrix similarity scores between each state's or county’s temporal correlation structure and the national pattern (Figure \ref{fig:fig4}). Highest alignment was found in Florida (RS = 1.00), Alabama (RS = 0.99), and Mississippi (RS = 0.99). Greatest divergence occurred in Montana (RS = 0.77), Utah (RS = 0.84), and South Dakota (RS = 0.86). County-level analysis revealed even broader variation, particularly in rural and Northwestern regions, suggesting localized documentation practices contribute to geospatial signal decay.

\subsection{Structural Drivers of Signal Decay}
To quantify how structural and sociodemographic factors influence diagnostic signal divergence, we regressed county-level RS scores on local characteristics (Table \ref{tab:table2}). We found that lower similarity to national coding patterns was associated with: (a) higher rural population share ($\beta$ = -0.19, p $<$ 0.05), (b) greater Medicaid eligibility among dementia patients ($\beta$ = -0.07), and (c) higher representation of Hispanic ($\beta$ = -0.16) and Black patients ($\beta$ = -0.13). In contrast, counties with higher proportions of dementia patients and those with more patients living with disabilities showed higher fidelity to national diagnostic patterns ($\beta$ = 1.27 and $\beta$ = 0.32, respectively). The model explained 38\% of the variance in diagnostic signal similarity (R$^2$ = 0.38), indicating that structural signal decay is partially (but not fully) explained by observable socioeconomic context.

\begin{table}[ht]
\centering
\begin{tabular}{l c}
\toprule
\textbf{Variables at the county level} & \textbf{Coefficient (SD)} \\
\midrule
Proportion of individuals with less than a high school education     & $-0.13\ (0.07)$ \\
Unemployment rate                                                   & $0.24\ (0.24)$ \\
Proportion of individuals residing in rural areas                   & $\mathbf{-0.19}^{*}(0.01)$ \\
Proportion of dementia patients eligible for Medicaid               & $\mathbf{-0.07}\ (0.03)$ \\
Proportion of dementia patients with disabilities                   & $\mathbf{0.32}\ (0.05)$ \\
Proportion of dementia patients                                     & $\mathbf{1.27}\ (0.10)$ \\
Proportion of Black dementia patients                               & $\mathbf{-0.13}\ (0.03)$ \\
Proportion of Hispanic dementia patients                            & $\mathbf{-0.16}\ (0.04)$ \\
Proportion of Asian dementia patients                               & $0.27\ (0.21)$ \\
\bottomrule
\end{tabular}
\begin{tablenotes}
\footnotesize
\item $^*$\textbf{Statistically significant} ($p<0.05$) coefficients are presented in bold.
\end{tablenotes}
\caption{\textbf{Estimated regression coefficients predicting the similarity of ICD-10 code utilization between counties and the nation.} The table presents the estimated coefficients and standard deviations (SD) obtained through the multi-variable linear regression with fixed effects. Each 10\% increase in a predictor variable corresponds to a difference in the county’s deviation from the national HUP, as measured by random skewers, equal to the value of the reported coefficient multiplied by 0.1.}
\label{tab:table2}
\end{table}

\subsection{Sensitivity Analyses}
Our results remained robust under alternative similarity metrics (mean absolute difference) and when repeating analyses using individual ICD-10 codes rather than grouped categories (Table \ref{tab:sens} in Supplementary Materials). These findings confirm that the observed signal decay patterns are not artifacts of coding aggregation or measurement method.

\section{Discussion}
The burden of dementia in the United States is deeply divergent, stratified by race, income, geography, and access to care \citep{kornblith2022association, katz2012age, mehta2017systematic, fitzpatrick2004incidence, plassman2007prevalence, moon2019dementia, mayeda2016inequalities, hayward2021importance, zhu2021sex, temkin2021racial, Fillenbaum2000, Husaini2003}. Yet beyond disparities in incidence and outcomes, this study reveals a more insidious layer of inequity: variability in how dementia is recorded in clinical data. We demonstrate that the informational quality of diagnosis codes – what we call diagnostic signal fidelity – is systematically shaped by structural and institutional factors. In regions with fewer resources, more racially minoritized populations, or limited access to specialty care, the fidelity of dementia documentation decays. This decay is not random; it follows lines of structural disadvantage.

By applying a novel temporal sequence mining approach (tSPM+) and cross-matrix similarity metrics, we quantified the architecture of dementia-related diagnostic codes across over 3,000 counties. Our findings show that in many regions, especially rural and under-resourced counties, dementia is more likely to be recorded using vague or non-specific codes, with temporal patterns diverging from national norms. This pattern, what we term diagnostic signal decay, reflects how local context alters the semantic and temporal structure of clinical documentation. Counties with high proportions of Black and Hispanic patients, higher Medicaid eligibility, or greater rurality showed significantly lower alignment with national diagnostic usage patterns, even after adjusting for covariates.

Urban counties showed the greatest alignment with national averages in dementia-related code utilization, potentially reflecting more integrated care between outpatient and hospital settings. In contrast, rural areas may lack local hospitals or specialist services, leading residents to seek care across county lines, which can fragment documentation and skew local coding patterns. While urban patients also have more provider options, stronger data-sharing networks may help standardize documentation. Future research incorporating outpatient data and care-seeking patterns could clarify how healthcare infrastructure shapes diagnostic coding variation.

Our observation of transitions from specific to non-specific diagnostic codes, particularly from Alzheimer’s disease to unspecified dementia, underscores a pattern of temporal signal decay, in which diagnostic specificity diminishes over time. In an idealized diagnostic trajectory, one might expect the process to unfold in the opposite direction: a general diagnosis made in a primary care setting would ideally be refined into a more specific subtype through subsequent specialist evaluation. However, our findings suggest that this refinement pathway is the exception rather than the norm. Instead, dementia is often first documented reactively, during acute hospitalizations, when functional decline becomes too severe to ignore, or when family members raise concerns, rather than proactively through structured diagnostic workups. The shift toward less specific codes over time may also reflect documentation fatigue or systemic factors, such as EHR workflows that promote code carry-forward rather than reassessment. Together, these dynamics contribute to a degradation of diagnostic clarity in longitudinal data and present challenges for both clinical management and secondary uses of administrative data.

These findings carry significant implications for health data science. In predictive modeling, diagnosis codes are often treated as static, reliable targets. But our results demonstrate that these codes may mean different things in different contexts; an Alzheimer’s diagnosis in Florida may reflect a well-resourced, specialist-evaluated case, while the same code in rural Montana may reflect diagnostic uncertainty or reimbursement-driven documentation. Such semantic instability can introduce a foundational vulnerability into AI models or any computational algorithm trained on claims or EHR data.

This decay is structured, not stochastic. It aligns with systemic inequities in healthcare access and infrastructure. Moreover, signal decay risks compromising the generalizability of predictive models across regions and populations. AI tools developed in high-fidelity zones, often large, urban, academic centers, may underperform when deployed in structurally distinct settings. This limitation is particularly relevant in ADRD, where avoiding diagnostic delays can be important for safety, care planning, and social support, and where emerging therapies may offer modest benefits when initiated earlier. By quantifying diagnostic signal fidelity, we move toward more context-aware and bias-sensitive modeling frameworks.

Our study also offers a methodological contribution: a replicable pipeline for assessing diagnostic signal fidelity in administrative data. The integration of tSPM+ with matrix similarity scoring enables the measurement of not only what codes are used, but also how they are sequenced and co-occur over time. This structural view of diagnosis coding may be particularly valuable in other complex conditions, such as long COVID, psychiatric disorders, and rheumatologic disorders, where semantic ambiguity is common and documentation can vary across settings and sub-populations.

We suggest that future work could explore the development of a Signal Fidelity Index (SFI), a composite measure that captures the clarity, consistency, and specificity of diagnostic code usage across healthcare settings. Such an index could provide a systematic way to assess the informational quality of clinical data and its susceptibility to signal decay. An SFI could serve as a contextual modifier in predictive modeling, a calibration tool for cross-site algorithm deployment, or a metric for auditing fairness across populations. Just as imaging studies report signal-to-noise ratios to quantify interpretive reliability, clinical AI may benefit from parallel measures that reflect the fidelity of its input data. While we do not propose a definitive formulation here, the concept offers a promising direction for quantifying and mitigating structural artifacts in real-world health data.

\subsection{Limitations}
Several limitations should be noted. First, our analysis is based on Medicare Part A claims, capturing inpatient hospitalizations but excluding outpatient, long-term care, and community-based services settings where dementia is often diagnosed or managed, which constrains the generalizability of our findings to the broader dementia population. Hospitalized patients differ systematically from those seen in outpatient settings: they are typically older, have more comorbid conditions, and are often at later stages of dementia when behavioral symptoms become more pronounced. Hospitalization itself may unmask previously unrecognized cognitive impairment due to the stress of acute illness or increased clinical attention, or it may precipitate stepwise decline or delirium – conditions that can be misclassified as dementia. Moreover, diagnostic codes in hospital settings are primarily assigned by professional coders rather than treating providers, which may further influence coding patterns and specificity. These factors should be considered when interpreting our findings and their relevance to outpatient or community-based populations.

Second, the use of county-level covariates introduces the possibility of ecological fallacy. While area-level indicators are essential for identifying contextual effects, they may obscure patient-level nuances. Third, while tSPM+ offers a robust tool for quantifying temporal code structure, it cannot distinguish between clinically warranted and administratively motivated transitions. Thus, some observed decay may reflect real clinical uncertainty rather than documentation failure.

Finally, while our model explained 38\% of the variation in local-to-national diagnostic pattern similarity, a substantial portion remains unexplained. This residual may represent unmeasured institutional factors, such as EHR configuration, local billing practices, training protocols, or limitations in our modeling approach. Without accounting for signal decay, computational models may propagate diagnostic artifacts, as if they were truths. Future work should expand to include health system-level metadata and chart-reviewed clinical validation to help disentangle variability in coding practices from true differences in disease prevalence and care-seeking behaviors, such as later presentation due to cultural, structural, or access-related factors.

\subsection{Conclusion}
Diagnostic codes are more than labels; they are the linguistic infrastructure of data-driven medicine. When that infrastructure degrades, through omission, vagueness, or fragmentation, it alters what we can know, predict, and act upon. This study shows that in the case of dementia, degradation in diagnostic coding is neither uniform nor inconsequential, though it often reflects documentation priorities tailored to patient care and billing rather than the nuanced needs of secondary data use. In both inpatient and outpatient settings, essential clinical details may reside in narrative notes rather than structured codes, underscoring the limits of relying solely on administrative data for population-level inferences. It is structurally patterned and consequential. To ensure that AI in healthcare is not only accurate but equitable, we must build tools that see not just the data, but the decay.

\bibliographystyle{unsrt}
\bibliography{biblio}

\begin{thebibliography}{10}

\bibitem{mmadumbu2025early}
Anthony~Chidubem Mmadumbu, Faisal Saeed, Fuad Ghaleb, and Sultan~Noman Qasem.
\newblock Early detection of alzheimer's disease using deep learning methods.
\newblock {\em Alzheimer's \& Dementia}, 21(5):e70175, 2025.

\bibitem{habbal2025harnessing}
Saadeddine Habbal, Maamoon Mian, Musa Imam, Jihane Tahiri, Adam Amor, and P~Hemachandra Reddy.
\newblock Harnessing artificial intelligence for transforming dementia care: Innovations in early detection and treatment.
\newblock {\em Brain Organoid and Systems Neuroscience Journal}, 2025.

\bibitem{hao2024early}
Jinkui Hao, William~R Kwapong, Ting Shen, Huazhu Fu, Yanwu Xu, Qinkang Lu, Shouyue Liu, Jiong Zhang, Yonghuai Liu, Yifan Zhao, et~al.
\newblock Early detection of dementia through retinal imaging and trustworthy ai.
\newblock {\em npj Digital Medicine}, 7(1):294, 2024.

\bibitem{sadeghian2024methods}
Roozbeh Sadeghian, Fasih Haider, Kathleen Fraser, Shinya Tasaki, and Graciela Muniz-Terrera.
\newblock Methods in artificial intelligence for dementia 2024, 2024.

\bibitem{kale2024ai}
Mayur~B Kale, Nitu~L Wankhede, Rupali~S Pawar, Suhas Ballal, Rohit Kumawat, Manish Goswami, Mohammad Khalid, Brijesh~G Taksande, Aman~B Upaganlawar, Milind~J Umekar, et~al.
\newblock Ai-driven innovations in alzheimer's disease: Integrating early diagnosis, personalized treatment, and prognostic modelling.
\newblock {\em Ageing Research Reviews}, page 102497, 2024.

\bibitem{bynum2024regional}
Julie~PW Bynum, Slim Benloucif, Jonathan Martindale, A~James O'Malley, and Matthew~A Davis.
\newblock Regional variation in diagnostic intensity of dementia among older us adults: an observational study.
\newblock {\em Alzheimer's \& Dementia}, 20(10):6755--6764, 2024.

\bibitem{dieleman2025drivers}
Joseph~L Dieleman, Maxwell Weil, Meera Beauchamp, Catherine Bisignano, Sawyer~W Crosby, Drew DeJarnatt, Haley Lescinsky, Ali~H Mokdad, Samuel Ostroff, Hilary Paul, et~al.
\newblock Drivers of variation in health care spending across us counties.
\newblock In {\em JAMA Health Forum}, volume~6, pages e245220--e245220. American Medical Association, 2025.

\bibitem{wennberg1999understanding}
John~E Wennberg.
\newblock Understanding geographic variations in health care delivery, 1999.

\bibitem{agniel2018biases}
Denis Agniel, Isaac~S Kohane, and Griffin~M Weber.
\newblock Biases in electronic health record data due to processes within the healthcare system: retrospective observational study.
\newblock {\em Bmj}, 361, 2018.

\bibitem{MedPAC2024}
{Medicare Payment Advisory Commission}.
\newblock June 2024 report to the congress: Medicare and the health care delivery system.
\newblock \url{https://www.medpac.gov/document/june-2024-report-to-the-congress-medicare-and-the-health-care-delivery-system/}, 2024.

\bibitem{li2021elucidating}
Linyan Li, George~F Chamoun, Nassib~G Chamoun, Daniel Sessler, Val{\'e}rie Gopinath, and Vikas Saini.
\newblock Elucidating the association between regional variation in diagnostic frequency with risk-adjusted mortality through analysis of claims data of medicare inpatients: a cross-sectional study.
\newblock {\em BMJ open}, 11(9):e054632, 2021.

\bibitem{mainor2019icd}
Alexander~J Mainor, Nancy~E Morden, Jeremy Smith, Stephanie Tomlin, and Jonathan Skinner.
\newblock Icd-10 coding will challenge researchers: caution and collaboration may reduce measurement error and improve comparability over time.
\newblock {\em Medical care}, 57(7):e42--e46, 2019.

\bibitem{xue2024ai}
Chonghua Xue, Sahana~S Kowshik, Diala Lteif, Shreyas Puducheri, Varuna~H Jasodanand, Olivia~T Zhou, Anika~S Walia, Osman~B Guney, J~Diana Zhang, Serena~T Pham, et~al.
\newblock Ai-based differential diagnosis of dementia etiologies on multimodal data.
\newblock {\em Nature Medicine}, 30(10):2977--2989, 2024.

\bibitem{lee2024robust}
Liz~Yuanxi Lee, Delshad Vaghari, Michael~C Burkhart, Peter Tino, Marcella Montagnese, Zhuoyu Li, Katharina Z{\"u}hlsdorff, Joseph Giorgio, Guy Williams, Eddie Chong, et~al.
\newblock Robust and interpretable ai-guided marker for early dementia prediction in real-world clinical settings.
\newblock {\em EClinicalMedicine}, 74, 2024.

\bibitem{kornblith2022association}
Erica Kornblith, Amber Bahorik, W~John Boscardin, Feng Xia, Deborah~E Barnes, and Kristine Yaffe.
\newblock Association of race and ethnicity with incidence of dementia among older adults.
\newblock {\em Jama}, 327(15):1488--1495, 2022.

\bibitem{katz2012age}
Mindy~J Katz, Richard~B Lipton, Charles~B Hall, Molly~E Zimmerman, Amy~E Sanders, Joe Verghese, Dennis~W Dickson, and Carol~A Derby.
\newblock Age-specific and sex-specific prevalence and incidence of mild cognitive impairment, dementia, and alzheimer dementia in blacks and whites: a report from the einstein aging study.
\newblock {\em Alzheimer Disease \& Associated Disorders}, 26(4):335--343, 2012.

\bibitem{mehta2017systematic}
Kala~M Mehta and Gwen~W Yeo.
\newblock Systematic review of dementia prevalence and incidence in united states race/ethnic populations.
\newblock {\em Alzheimer's \& Dementia}, 13(1):72--83, 2017.

\bibitem{fitzpatrick2004incidence}
Annette~L Fitzpatrick, Lewis~H Kuller, Diane~G Ives, Oscar~L Lopez, William Jagust, John~CS Breitner, Beverly Jones, Constantine Lyketsos, and Corinne Dulberg.
\newblock Incidence and prevalence of dementia in the cardiovascular health study.
\newblock {\em Journal of the American Geriatrics Society}, 52(2):195--204, 2004.

\bibitem{plassman2007prevalence}
Brenda~L Plassman, Kenneth~M Langa, Gwenith~G Fisher, Steven~G Heeringa, David~R Weir, Mary~Beth Ofstedal, James~R Burke, Michael~D Hurd, Guy~G Potter, Willard~L Rodgers, et~al.
\newblock Prevalence of dementia in the united states: the aging, demographics, and memory study.
\newblock {\em Neuroepidemiology}, 29(1-2):125--132, 2007.

\bibitem{mccarthy2022validation}
Ellen~P McCarthy, Chiang-Hua Chang, Nicholas Tilton, Mohammed~U Kabeto, Kenneth~M Langa, and Julie~PW Bynum.
\newblock Validation of claims algorithms to identify alzheimer’s disease and related dementias.
\newblock {\em The Journals of Gerontology: Series A}, 77(6):1261--1271, 2022.

\bibitem{ccw}
Chronic Conditions~Data Warehouse.
\newblock Home | chronic conditions data warehouse, 2024.

\bibitem{moon2019dementia}
Heehyul Moon, Adrian~NS Badana, So-Yeon Hwang, Jeanelle~S Sears, and William~E Haley.
\newblock Dementia prevalence in older adults: Variation by race/ethnicity and immigrant status.
\newblock {\em The American Journal of Geriatric Psychiatry}, 27(3):241--250, 2019.

\bibitem{mayeda2016inequalities}
Elizabeth~Rose Mayeda, M~Maria Glymour, Charles~P Quesenberry, and Rachel~A Whitmer.
\newblock Inequalities in dementia incidence between six racial and ethnic groups over 14 years.
\newblock {\em Alzheimer's \& Dementia}, 12(3):216--224, 2016.

\bibitem{hayward2021importance}
Mark~D Hayward, Mateo~P Farina, Yuan~S Zhang, Jung~Ki Kim, and Eileen~M Crimmins.
\newblock The importance of improving educational attainment for dementia prevalence trends from 2000 to 2014, among older non-hispanic black and white americans.
\newblock {\em The Journals of Gerontology: Series B}, 76(9):1870--1879, 2021.

\bibitem{ornstein2018medicare}
Katherine~A Ornstein, Carolyn~W Zhu, Evan Bollens-Lund, Melissa~D Aldridge, Howard Andrews, Nicole Schupf, and Yaakov Stern.
\newblock Medicare expenditures and health care utilization in a multiethnic community-based population with dementia from incidence to death.
\newblock {\em Alzheimer Disease \& Associated Disorders}, 32(4):320--325, 2018.

\bibitem{gorges2019national}
Rebecca~J Gorges, Prachi Sanghavi, and R~Tamara Konetzka.
\newblock A national examination of long-term care setting, outcomes, and disparities among elderly dual eligibles.
\newblock {\em Health Affairs}, 38(7):1110--1118, 2019.

\bibitem{gessert2006rural}
Charles~E Gessert, Irina~V Haller, Robert~L Kane, and Howard Degenholtz.
\newblock Rural--urban differences in medical care for nursing home residents with severe dementia at the end of life.
\newblock {\em Journal of the American Geriatrics Society}, 54(8):1199--1205, 2006.

\bibitem{estiri2020transitive}
Hossein Estiri, Sebastien Vasey, and Shawn~N Murphy.
\newblock Transitive sequential pattern mining for discrete clinical data.
\newblock In {\em Artificial Intelligence in Medicine: 18th International Conference on Artificial Intelligence in Medicine, AIME 2020, Minneapolis, MN, USA, August 25--28, 2020, Proceedings 18}, pages 414--424. Springer, 2020.

\bibitem{hugel2023tspm+}
Jonas H{\"u}gel, Ulrich Sax, Shawn~N Murphy, and Hossein Estiri.
\newblock tspm+; a high-performance algorithm for mining transitive sequential patterns from clinical data.
\newblock {\em arXiv preprint arXiv:2309.05671}, 2023.

\bibitem{hugel2023tspmplus}
J.~Hügel.
\newblock tspmplus\_r: R package for the tspmplus algorithm.
\newblock \url{https://github.com/jhuegel/tSPMPlus_R}, 2023.
\newblock GitHub repository.

\bibitem{zhu2021sex}
Yingying Zhu, Yi~Chen, Eileen~M Crimmins, and Julie~M Zissimopoulos.
\newblock Sex, race, and age differences in prevalence of dementia in medicare claims and survey data.
\newblock {\em The Journals of Gerontology: Series B}, 76(3):596--606, 2021.

\bibitem{dartmouth1996atlas}
{Dartmouth Medical School Center for the Evaluative Clinical Sciences}.
\newblock {\em The Dartmouth Atlas of Health Care}.
\newblock Health Forum Publishing Company, 1996.

\bibitem{smith2011dartmouth}
Richard Smith.
\newblock Dartmouth atlas of health care, 2011.

\bibitem{badinski2023geographic}
Ivan Badinski, Amy Finkelstein, Matthew Gentzkow, and Peter Hull.
\newblock Geographic variation in healthcare utilization: The role of physicians.
\newblock Technical report, National Bureau of Economic Research, 2023.

\bibitem{marshall2022diagnosis}
Trisha~L Marshall, Philip~A Hagedorn, Courtney Sump, Chelsey Miller, Matthew Fenchel, Dane Warner, Anna~J Ipsaro, Peter O’Day, Todd Lingren, and Patrick~W Brady.
\newblock Diagnosis code and health care utilization patterns associated with diagnostic uncertainty.
\newblock {\em Hospital Pediatrics}, 12(12):1066--1072, 2022.

\bibitem{geruso2020upcoding}
Michael Geruso and Timothy Layton.
\newblock Upcoding: evidence from medicare on squishy risk adjustment.
\newblock {\em Journal of Political Economy}, 128(3):984--1026, 2020.

\bibitem{crespin2024upcoding}
Daniel Crespin, Michael Dworsky, Jonathan Levin, Teague Ruder, and Christopher~M Whaley.
\newblock Upcoding linked to up to two-thirds of growth in highest-intensity hospital discharges in 5 states, 2011--19: Study examines hospital patient discharges with highest-intensity charge codes.
\newblock {\em Health Affairs}, 43(12):1619--1627, 2024.

\bibitem{carter1990much}
Grace~M Carter, Joseph~P Newhouse, and Daniel~A Relles.
\newblock How much change in the case mix index is drg creep?
\newblock {\em Journal of health economics}, 9(4):411--428, 1990.

\bibitem{drgcreep1981}
Drg creep.
\newblock {\em The New England Journal of Medicine}, 305:961--962, 1981.

\bibitem{Manson2021}
Steven Manson, Jonathan Schroeder, David Van~Riper, Tracy Kugler, and Steven Ruggles.
\newblock National historical geographic information system: Version 16.0, 2021.

\bibitem{cheverud2007research}
James~M Cheverud and Gabriel Marroig.
\newblock Research article comparing covariance matrices: random skewers method compared to the common principal components model.
\newblock {\em Genetics and Molecular Biology}, 30:461--469, 2007.

\bibitem{rohlf2017method}
F~James Rohlf.
\newblock The method of random skewers.
\newblock {\em Evolutionary Biology}, 44(4):542--550, 2017.

\bibitem{temkin2021racial}
Helena Temkin-Greener, Di~Yan, Sijiu Wang, and Shubing Cai.
\newblock Racial disparity in end-of-life hospitalizations among nursing home residents with dementia.
\newblock {\em Journal of the American Geriatrics Society}, 69(7):1877--1886, 2021.

\bibitem{Fillenbaum2000}
G.~Fillenbaum, A.~Heyman, B.~Peterson, C.~Pieper, and A.L. Weiman.
\newblock Frequency and duration of hospitalization of patients with ad based on medicare data: Cerad xx.
\newblock {\em Neurology}, 54(3):740–740, February 2000.

\bibitem{Husaini2003}
Baqar~A. Husaini, Darren~E. Sherkat, Majaz Moonis, Robert Levine, Charles Holzer, and Van~A. Cain.
\newblock Racial differences in the diagnosis of dementia and in its effects on the use and costs of health care services.
\newblock {\em Psychiatric Services}, 54(1):92–96, January 2003.

\end{thebibliography}

\newpage
\section{Supplementary Materials}

\renewcommand{\thesection}{S\arabic{section}}
\renewcommand{\thefigure}{S\arabic{figure}}
\renewcommand{\thetable}{S\arabic{table}}
\renewcommand{\theequation}{S\arabic{equation}}

\setcounter{section}{0}
\setcounter{figure}{0}
\setcounter{table}{0}
\setcounter{equation}{0}
\begin{figure}[h]
    \centering
    \includegraphics[width=0.9\linewidth]{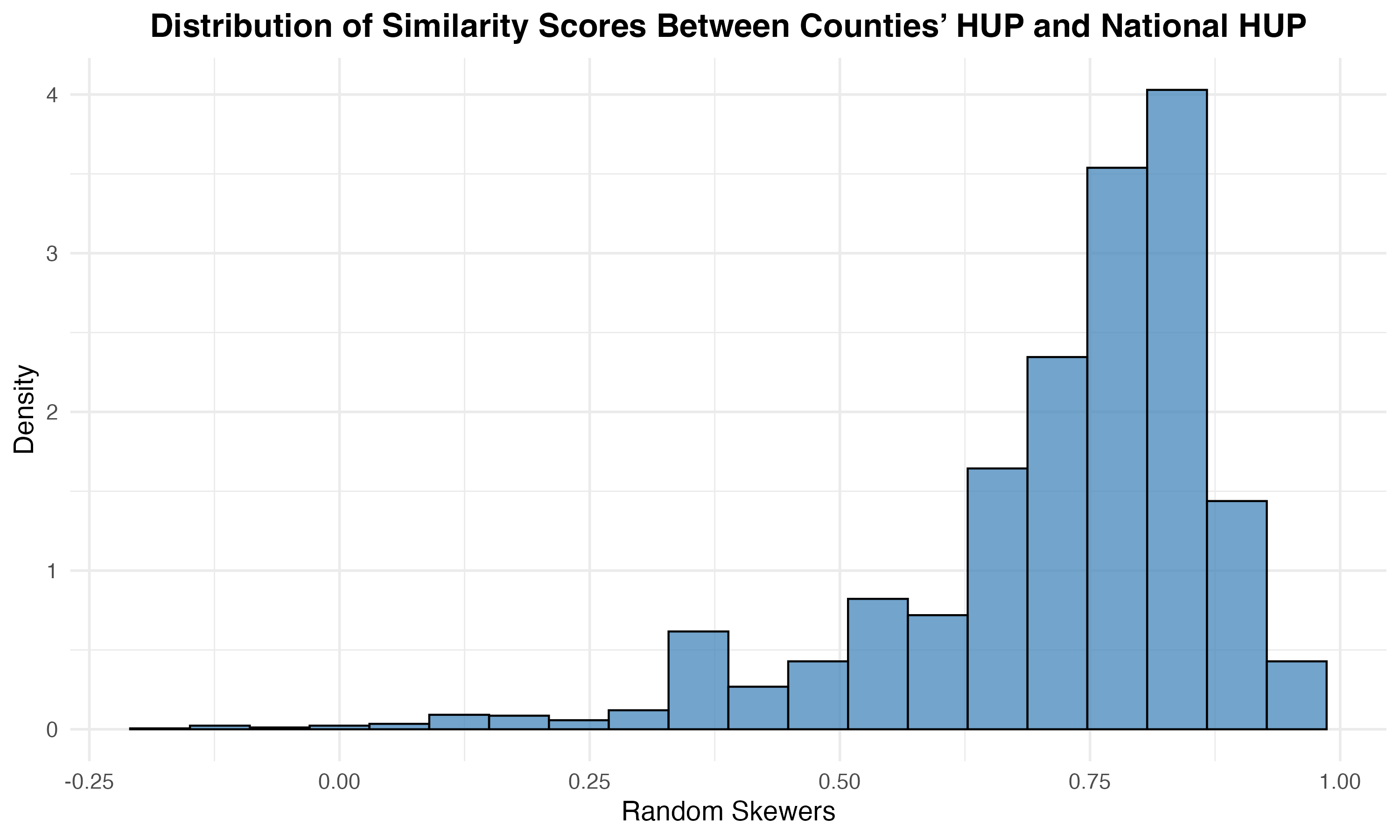}
    \caption{\textbf{Distribution of random skewers metric.} Distribution of the similarity score between each county’s dementia diagnostic pattern and the national reference for the period 2016-2018, as computed by random skewers.}
    \label{fig:figS1}
\end{figure}

\begin{table}[ht]
\centering
\small
\begin{tabular}{@{}p{2.3cm} p{3.7cm} p{9cm}@{}}
\toprule
\textbf{ICD-10 code} & \textbf{Diagnostic category} & \textbf{Description} \\
\midrule
G300  & Alzheimer's disease     & Alzheimer's disease with early onset \\
G301  & Alzheimer's disease     & Alzheimer's disease with late onset \\
G308  & Alzheimer's disease     & Other Alzheimer's disease \\
G309  & Alzheimer's disease     & Alzheimer's disease, unspecified \\
G3183 & Neurocognitive disorder & Neurocognitive disorder with Lewy bodies \\
G3185 & Neurocognitive disorder & Corticobasal degeneration \\
F0280 & Non-specific dementia   & Dementia in other diseases classified elsewhere, unspecified severity, without behavioral disturbance, psychotic disturbance, mood disturbance, and anxiety \\
F0281 & Non-specific dementia   & Dementia in other diseases classified elsewhere, unspecified severity, with behavioral disturbance \\
F0390 & Non-specific dementia   & Unspecified dementia, unspecified severity, without behavioral disturbance, psychotic disturbance, mood disturbance, and anxiety \\
F0391 & Non-specific dementia   & Unspecified dementia, unspecified severity, with behavioral disturbance \\
G3109 & Non-specific dementia   & Other frontotemporal neurocognitive disorder \\
G311  & Non-specific dementia   & Senile degeneration of brain, not elsewhere classified \\
G3189 & Non-specific dementia   & Other specified degenerative diseases of nervous system \\
G319  & Non-specific dementia   & Degenerative disease of nervous system, unspecified \\
G3101 & Pick's disease          & Pick's disease \\
F0150 & Vascular dementia       & Vascular dementia, unspecified severity, without behavioral disturbance, psychotic disturbance, mood disturbance, and anxiety \\
F0151 & Vascular dementia       & Vascular dementia, unspecified severity, with behavioral disturbance \\
\bottomrule
\end{tabular}
\caption{\textbf{Codes for dementia-qualifying hospitalization identification and corresponding diagnostic categories.} The table presents the list of the ICD-10 codes considered for the analysis, their diagnostic category, and their description.}
\label{tab:codes}
\end{table}

\begin{table}[ht]
\centering

\begin{tabular}{@{}p{6cm}ccc@{}}
\toprule
\textbf{} & \textbf{5 categories} & \textbf{5 categories} & \textbf{17 ICD-10 code} \\
\textbf{Variables} & \textbf{regression (RS)} & \textbf{regression (1-MAD)} & \textbf{regression (RS)} \\
\midrule
Rate of individuals with less than a high school education          & $-0.13\ (0.07)$ & $-0.03\ (0.01)$ & $\mathbf{-0.19}\ (0.05)$ \\
Unemployment rate                                                   & $0.24\ (0.24)$  & $0.05\ (0.04)$  & $0.27\ (0.20)$ \\
Rate of residents residing in rural areas                           & $\mathbf{-0.19}\ (0.01)$ & $\mathbf{-0.04}\ (0.01)$ & $\mathbf{-0.25}\ (0.01)$ \\
Rate of dementia patients eligible for Medicaid                     & $\mathbf{-0.07}\ (0.03)$ & $\mathbf{-0.01}\ (0.00)$ & $\mathbf{-0.10}\ (0.03)$ \\
Rate of dementia patients with disabilities                         & $\mathbf{0.32}\ (0.05)$  & $\mathbf{0.05}\ (0.01)$  & $\mathbf{0.33}\ (0.05)$ \\
Rate of dementia patients                                           & $\mathbf{1.27}\ (0.10)$  & $\mathbf{0.19}\ (0.02)$  & $\mathbf{1.55}\ (0.08)$ \\
Rate of Black dementia patients                                     & $\mathbf{-0.13}\ (0.03)$ & $\mathbf{-0.11}\ (0.00)$ & $\mathbf{-0.12}\ (0.02)$ \\
Rate of Hispanic dementia patients                                  & $\mathbf{-0.16}\ (0.04)$ & $\mathbf{-0.03}\ (0.01)$ & $\mathbf{-0.14}\ (0.03)$ \\
Rate of Asian dementia patients                                     & $0.27\ (0.21)$  & $0.04\ (0.04)$  & $\mathbf{0.55}\ (0.17)$ \\
\bottomrule
\end{tabular}
\caption{\textbf{Estimated regression coefficients predicting the similarity of ICD-10 code utilization between counties and the nation.} The table presents the estimated coefficients and standard deviations (SD) for the main and sensitivity analyses obtained through the multi-variable linear regression with fixed effects. Each 10\% increase in a predictor variable corresponds to a difference in the county’s deviation from the national HUP equal to the value of the reported coefficient multiplied by 0.1. The significant coefficients are presented in bold. }
\label{tab:sens}
\end{table}

\end{document}